\documentclass[a4paper,12pt]{article}
\usepackage[margin=2.5cm]{geometry}
\usepackage{setspace}
\usepackage{indentfirst}
\usepackage{footmisc}
\usepackage{amsmath}
\usepackage{amssymb}
\usepackage[nosort]{cite}
\usepackage[svgnames]{xcolor}
\usepackage[colorlinks=true,
            allcolors=.,
            bookmarksnumbered=true,
            pdfpagemode=UseNone,
            pdfstartview=FitH]{hyperref}

\newcommand{\dif}{\text{d}}

\numberwithin{equation}{section}
\makeatletter
\renewcommand\section{\@startsection {section}{1}{\z@}%
	{-3.5ex \@plus -1ex \@minus -.2ex}%
	{2.3ex \@plus.2ex}%
	{\normalfont\sffamily\bfseries}}
\renewcommand\subsection{\@startsection{subsection}{2}{\z@}%
	{-3.25ex\@plus -1ex \@minus -.2ex}%
	{1.5ex \@plus .2ex}%
	{\normalfont\sffamily\slshape}}
\makeatother

\begin{document}

\thispagestyle{empty}
\vbox{}
\vspace{2cm}

\begin{center}
  {\sffamily\LARGE{Multi-centered rotating black holes in \\[2mm] Kaluza--Klein theory
  }}\\[16mm]
  {\sffamily Edward Teo~~and~~Thomas Wan}
  \\[6mm]
    {\sffamily\slshape\selectfont
      Department of Physics, National University of Singapore, 
      Singapore
    }\\[15mm]
\end{center}
\vspace{2cm}
	
\centerline{\sffamily\bfseries Abstract}
\bigskip
\noindent  
The most general charged and rotating black hole in Kaluza--Klein theory is known to be described by the Rasheed--Larsen solution. When the under-rotating extremal limit of this solution is taken, it falls into a general class of solutions of Kaluza--Klein theory found by Cl\'ement, and is specified by two harmonic functions on a three-dimensional flat base space. We use this fact to generalise the single extremal black hole solution to one describing an arbitrary superposition of such black holes. These black holes carry non-zero electric and magnetic charges, which we set to be equal for simplicity, and are in general rotating with parallel or anti-parallel spin vectors. It is checked that the space-time outside the black holes is free of pathologies such as naked singularities and closed time-like curves.

\newpage

\section{Introduction}

Exact solutions of Einstein's field equations describing two or more black holes are of great interest, as they provide a window into the nature of the interactions between black holes. However, due to the non-linearity of the field equations, such solutions are not easy to come by. An example is the Israel--Khan solution \cite{Israel:1964}, which describes an arbitrary number of Schwarzschild black holes placed along a symmetry axis. The black holes are kept in static equilibrium by conical singularities stretching between them, which are necessarily present to counterbalance their gravitational attraction.

If we require the space-time to be free of conical singularities and other defects outside the black holes, then there must be another force present---such as electromagnetism---to counterbalance gravity. An example is the Majumdar--Papapetrou solution \cite{Majumdar:1947,Papapetrou:1947} of Einstein--Maxwell theory, which describes an arbitrary superposition of Reissner--Nord\-str\"om black holes. In this case, the black holes have to be extremally charged, so that the electrostatic repulsion between them exactly cancels out their gravitational attraction. Furthermore, the black holes can be located anywhere in the space-time, unlike those in the Israel--Khan solution.

It is natural to wonder if there exists a rotating generalisation of the Majumdar--Papapetrou solution. Such a solution was found by  Israel and Wilson \cite{Israel:1972} and Perj\'es \cite{Perjes:1971}, but it turns out that this solution describes a superposition of naked singularities rather than black holes. It appears that a balanced superposition of rotating black holes is not possible in Einstein--Maxwell theory.

The purpose of this paper is to point out that a balanced superposition of rotating black holes is possible, if we consider instead Kaluza--Klein theory \cite{Kaluza:1921,Klein:1926}. This theory arises from the dimensional reduction of Einstein gravity in five space-time dimensions, and was originally conceived as a way to unify the gravitational and electromagnetic forces. However, it also gives rise to a new massless scalar field, known as the dilaton, which has not been observed. Nevertheless, the Kaluza--Klein method of dimensional reduction has found applications in modern theories such as string theory.

The black holes of Kaluza--Klein theory have been extensively studied over the years, beginning with the pioneering work of \cite{Leutwyler:1960,Dobiasch:1982,Chodos:1982,Pollard:1983,Gibbons:1986}. The most general black hole solution of Kaluza--Klein theory was found by Rasheed \cite{Rasheed:1995}, and independently by Larsen \cite{Larsen:1999,Larsen:2000}.\footnote{It was also independently found in \cite{Matos:1996}.} This black hole is specified by four parameters: mass, angular momentum, electric charge and magnetic charge. Its various properties were studied in detail in \cite{Rasheed:1995,Larsen:1999,Larsen:2000}. In particular, it was found that there are two different ways to take the extremal limit of this black hole.

It is one of these extremal limits, known as the under-rotating limit, that would be of interest to us. In this limit, the electric and magnetic charges are both non-zero, and the angular momentum is free to take any value in a certain range. This results in an extremally charged, rotating black hole solution that takes a very simple mathematical form. In fact, it falls into a class of solutions of Kaluza--Klein theory found by Cl\'ement \cite{Clement:1985,Clement:1986}, and is specified by a set of two harmonic functions on a three-dimensional flat base space, which have poles at the location of the black hole. By superposing multiple copies of these harmonic functions, we are able to construct a multi-centered generalisation of this black hole.

We note that the use of harmonic functions on a flat base space to construct multi-centered black hole solutions is common in supergravity theories, with the balance of forces between the black holes attributed to the presence of supersymmetry. Supersymmetric solutions describing multi-centered rotating black holes in four dimensions have been found in, e.g., \cite{Denef:2000,Denef:2000a,Bates:2003}. Even the Israel--Wilson--Perj\'es solution---though not strictly a black hole solution---can be embedded in $N=2$ supergravity. It is therefore somewhat unusual that the multi-centered rotating black hole solution constructed in this paper is {\it not\/} supersymmetric \cite{Larsen:1999,Larsen:2000}, even though the forces between the black holes balance out.

This paper is organised as follows: In Sec.~\ref{KK sec}, we review  elements of Kaluza--Klein theory, in particular the formalism of Maison \cite{Maison:1979}, as well as the general class of solutions found by Cl\'ement \cite{Clement:1985,Clement:1986}. In Sec.~\ref{Rotating black hole sec}, we present the Rasheed--Larsen black hole solution, and point out how one of its extremal limits falls into the class of solutions found by Cl\'ement. We then show in Sec.~\ref{Rotating multi black hole sec} how this single black hole solution, with equal electric and magnetic charges, can be generalised to one describing an arbitrary superposition of such black holes. Various properties of this multi-centered rotating black hole solution are studied in Sec.~\ref{Properties sec}. In Sec.~\ref{Special case sec}, we present the explicit solution for two black holes lying along a symmetry axis. The paper concludes with a summary of the results obtained and some avenues for future work.

\section{Kaluza--Klein theory}
\label{KK sec}

The starting point of Kaluza--Klein theory is Einstein gravity in five space-time dimensions, with the assumption that the metric is invariant under translations of the extra space dimension $x^5$. Moreover, it is assumed that $x^5$ is periodic with period $2\pi R_{\rm KK}$. In this case, the five-dimensional metric can be written in the form
\begin{align}
\label{5D metric}
\dif s_{(5)}^2={\rm e}^{-\frac{2\phi}{\sqrt3}}\big(\dif x^5+A_\mu\dif x^\mu\big)^2+{\rm e}^{\frac{\phi}{\sqrt3}}g_{\mu\nu}\dif x^\mu\dif x^\nu,
\end{align}
where $\mu,\nu=0,\dots,3$, and the functions $\phi$, $A_\mu$ and $g_{\mu\nu}$ depend only on the coordinates $x^\mu$. The five-dimensional Einstein--Hilbert action then reduces to the four-dimensional action
\begin{align}
2\pi R_{\rm KK}\int\dif^4 x\sqrt{-g}\left(R-\frac{1}{2}\partial_\mu\phi\partial^\mu\phi-\frac{1}{4}{\rm e}^{-\sqrt{3}\phi}F_{\mu\nu}F^{\mu\nu}\right),
\end{align}
where $R$ is the four-dimensional Ricci scalar, $g=\det g_{\mu\nu}$, and $F_{\mu\nu}=\partial_\mu A_\nu-\partial_\nu A_\mu$. This shows that Kaluza--Klein theory can be regarded as a type of Einstein--Maxwell-dilaton theory in four dimensions. The corresponding field equations are
\begin{subequations}
\label{field equations}
\begin{gather}
R_{\mu\nu}=\frac{1}{2}\partial_\mu\phi\partial_\nu\phi+\frac{1}{2}{\rm e}^{-\sqrt{3}\phi}\left(F_{\mu\rho}F_\nu{}^\rho-\frac{1}{4}g_{\mu\nu}F_{\rho\sigma}F^{\rho\sigma}\right),\\
\nabla_\mu\big({\rm e}^{-\sqrt{3}\phi}F^{\mu\nu}\big)=0\,,\\
\nabla_\mu\nabla^\mu\phi=-\frac{\sqrt3}{4}{\rm e}^{-\sqrt{3}\phi}F_{\mu\nu}F^{\mu\nu}.
\end{gather}
\end{subequations}

The field equations (\ref{field equations}) are difficult to solve without further assumptions. At this stage, it is usual to assume that the solution is stationary and possesses some spatial symmetry. However, in this paper, we only assume the former. As shown by Maison \cite{Maison:1979}, this is sufficient to enable the field equations to be written in an elegant, simplified form. We now briefly describe this formalism. 

If we further require that the five-dimensional metric (\ref{5D metric}) is invariant under translations of the time coordinate $x^0=t$, it can be written in the form
\begin{align}
\label{5D}
\dif s_{(5)}^2=\lambda_{ab}\big(\dif x^a +\omega^a{}_i\dif x^i\big)\big(\dif x^b +\omega^b{}_j \dif x^j\big)+\frac{1}{\tau}h_{ij}\dif x^i \dif x^j,
\end{align}
where $\tau\equiv-\det\lambda_{ab}$. Here, $a,b=0,5$, while $i,j=1,2,3$. The functions $\lambda_{ab}$, $\omega^a{}_i$ and $h_{ij}$ depend only on the three spatial coordinates $x^i$. From these functions, we can construct the symmetric, unimodular $3\times3$ matrix:
\begin{align}
\label{chi}
\chi=\begin{pmatrix}
\lambda_{ab}-\frac{1}{\tau}V_aV_b & \frac{1}{\tau}V_a \\
\frac{1}{\tau}V_b  & -\frac{1}{\tau}
\end{pmatrix},
\end{align}
where the so-called twist potentials $V_a$ are given by
\begin{align}
\label{twist}
V_{a,i}=\tau\lambda_{ab}\varepsilon_i{}^{jk}\omega^b{}_{j,k}\,.
\end{align}
Here, the spatial indices $i,j,\dots$ are raised and lowered using the metric $h_{ij}$, and $,i$ denotes the partial derivative  $\frac{\partial}{\partial x^i}$. With the definitions (\ref{chi}) and (\ref{twist}), the five-dimensional Einstein equation can be written in the simple and suggestive form:
\begin{subequations}
\label{field equations 2}
\begin{gather}
\label{field equations 2a}
R_{ij}=\frac{1}{4}\,{\rm Tr}\,\big(\chi^{-1}\chi_{,i}\chi^{-1}\chi_{,j}\big)\,,\\
\label{field equations 2b}
\big(\chi^{-1}\chi^{,i}\big)_{;i}=0\,,
\end{gather}
\end{subequations}
where $R_{ij}$ is the three-dimensional Ricci tensor corresponding to $h_{ij}$, and $;i$ denotes the covariant derivative with respect to $h_{ij}$.

We note that the field equations (\ref{field equations 2}) can be obtained from the three-dimensional $\sigma$-model action
\begin{align}
\int\dif^3 x\sqrt{h}\left[{}^{(3)}R-\frac{1}{4}\,{\rm Tr}\,\big(\chi^{-1}\chi_{,i}\chi^{-1}\chi^{,i}\big)\right]. 
\end{align}
This action has a manifest $SL(3,\mathbb{R})$ symmetry \cite{Maison:1979}, which can be used to generate new solutions from known ones. However, we do not use this approach here; instead, we construct solutions to the field equations (\ref{field equations 2}) directly. 

In this paper, we are specifically interested in the case of a {\it flat\/} three-dimensional metric $h_{ij}$. A very general class of solutions to (\ref{field equations 2}) was found by Cl\'ement \cite{Clement:1985,Clement:1986} for this case. These solutions can be written in the form 
\begin{align}
\label{clement}
\chi=\eta{\rm e}^{Af}{\rm e}^{A^2g},
\end{align}
where $\eta$ and $A$ are constant $3\times3$ matrices, and $f$ and $g$ are harmonic functions of $x^i$. If $f$ and $g$ vanish at infinity, then $\eta$ is given by
\begin{align}
\label{eta}
\eta=\begin{pmatrix}
-1 & 0 & 0 \\
0 & 1 & 0 \\
0 & 0 & -1
\end{pmatrix},
\end{align}
which will ensure that the four-dimensional space-time is asymptotically Minkowski. Since $\chi$ is symmetric and unimodular, we require that $A^{\rm T}=\eta A\eta$, and that $A$ and $A^2$ are both traceless. Furthermore, we require that $A^3$ is also traceless, so that (\ref{field equations 2a}) is satisfied.

\section{Rotating black hole solution}
\label{Rotating black hole sec}

The most general rotating black hole solution of Kaluza--Klein theory was first found by Rasheed \cite{Rasheed:1995}. It was constructed by performing a set of SO(1,2) transformations on the Kerr solution, using the formalism of Maison \cite{Maison:1979} reviewed in Sec.~\ref{KK sec}. This black hole solution can be parameterised by its mass $M$, angular momentum $J$, electric charge $Q$, and magnetic charge $P$. 

Now, the three-dimensional metric $h_{ij}$ of this solution is only flat when a certain extremal limit is taken, known as the under-rotating limit. The resulting class of extremal solutions satisfies the conditions \cite{Rasheed:1995}
\begin{align}
\Big(\frac{P}{M}\Big)^{\frac{2}{3}}+\Big(\frac{Q}{M}\Big)^{\frac{2}{3}}=2^{\frac{2}{3}}, \qquad |J|<|PQ|\,.
\end{align}
In the phase diagram Fig.~2 of \cite{Rasheed:1995}, it forms a vertical wall (labelled by W) which joins up to another class of extremal solutions along the $|J|=|PQ|$ curve.

The under-rotating class of extremal solutions is perhaps most compactly expressed in the form found by Larsen \cite{Larsen:1999}. It can be recovered from the general solution in \cite{Larsen:1999}\footnote{Note that there is a sign error in the general solution given in \cite{Larsen:1999}. Here, we use the corrected form of the solution given in \cite{Emparan:2007,Azeyanagi:2008}.} by taking the limit $m,a\rightarrow0$, such that $\frac{a}{m}\rightarrow j$ where $j$ is a finite constant satisfying $|j|<1$. We then obtain the five-dimensional metric:
\begin{align}
\label{larsen}
\dif s_{(5)}^2&=\frac{H_2}{H_1}\bigg\{\dif x^5- \left[ 2 \left( r+\frac{p}{2} \right) -{
pj\cos\theta} \right] \frac{Q}{H_2}\,\dif t\cr
&\quad\qquad- \left[ 2H_2\cos\theta
 -q \left( r+{\frac {pq}{p+q}} \right) j\sin^2 \theta
 \right]\frac{P}{H_2}\, \dif\varphi \bigg\} ^{2}\cr
&\quad-\frac{{r}^{2}}{H_2}
 \left( \dif t+{\frac {2jPQ \sin^2\theta \,
 \dif\varphi}{r}} \right) ^{2}+H_1
 \left( {\frac {{\dif r}^{2}}{{r}^{2}}}+{\dif\theta}^{2}+
\sin^2 \theta\,{\dif\varphi}^{2} \right),
\end{align}
where
\begin{subequations}
\begin{align}
H_1&\equiv{r}^{2}+rp+{\frac {{p}^{2}q \left( 1+j\cos\theta
  \right) }{2(p+q)}}\,,\\
H_2&\equiv{r}^{2}+rq+{\frac {p{q}^{2} \left( 1-j\cos \theta
   \right) }{2(p+q)}}\,.
\end{align}
\end{subequations}
Here, $(r,\theta,\varphi)$ are the standard spherical polar coordinates. The parameters $p$ and $q$ satisfy $p,q>0$, and are related to the magnetic and electric charges by
\begin{align}
P^2=\frac {{p}^{3}}{4(p+q)}\,,\qquad Q^2=\frac {{q}^3}{4(p+q)}\,.
\end{align}
For simplicity, we henceforth assume that $P$ and $Q$ are positive. We also note that the mass and angular momentum of the black hole are given by
\begin{align}
M=\frac{p+q}{4}\,,\qquad J= jPQ\,.
\end{align}

It can be checked, after some calculation, that this solution can be written in the form (\ref{clement}), with
\begin{align}
A=4\begin{pmatrix}
-\frac{q}{p+q} & \big(\frac{q}{p+q}\big)^\frac{3}{2} & 0 \\
-\big(\frac{q}{p+q}\big)^\frac{3}{2} & -\frac{p-q}{p+q} & \big(\frac{p}{p+q}\big)^\frac{3}{2} \\
0 & -\big(\frac{p}{p+q}\big)^\frac{3}{2} & \frac{p}{p+q} 
\end{pmatrix}.
\end{align}
Note that $A^3=0$, so that the exponentials in (\ref{clement}) have series expansions that are actually finite. The harmonic functions $f$ and $g$ are given by
\begin{align}
f=\frac{M}{r}\,,\qquad g=\frac{JM^2}{2PQ}\frac{\cos\theta}{r^2}\,.
\end{align}

A particularly simple case of this solution arises when $p=q$, i.e., when the magnetic and electric charges are equal: $P=Q=\frac{M}{\sqrt2}$. In this case, the matrix $A$ becomes
\begin{align}
\label{A}
A=\begin{pmatrix}
-2 & \sqrt{2} & 0 \\
-\sqrt{2} & 0 & \sqrt{2} \\
0 & -\sqrt{2} &  2
\end{pmatrix},
\end{align}
while the harmonic functions become
\begin{align}
\label{harmonic}
f=\frac{M}{r}\,,\qquad g=\frac{J\cos\theta}{r^2}\,.
\end{align}
The solution (\ref{larsen}) then becomes
\begin{align}
\dif s_{(5)}^2&=\frac{H_2}{H_1}\bigg\{\dif x^5- \sqrt{2}\,H_2^{-1}\left[M(r+M) -{
2J\cos\theta} \right] \dif t\cr
&\quad\qquad- \sqrt{2}\left[M \cos\theta
 -2J H_2^{-1}(r+M)\sin^2 \theta
 \right] \dif\varphi \bigg\} ^{2}\cr
&\quad-\frac{{r}^{2}}{H_2}
 \left( \dif t+{\frac {2J \sin^2\theta \,
 \dif\varphi}{r}} \right) ^{2}+H_1
 \left( {\frac {{\dif r}^{2}}{{r}^{2}}}+{\dif\theta}^{2}+
\sin^2 \theta\,{\dif\varphi}^{2} \right),
\end{align}
where now $H_{1,2}=(r+M)^{2}\pm2J\cos\theta$. The four-dimensional form of this solution, after substituting in for $H_{1,2}$ and redefining $r\rightarrow r-M$, is given by 
\begin{subequations}
\label{single black hole solution}
\begin{align}
\label{single black hole metric}
\dif s_{(4)}^2&=-\frac{(r-M)^2}{\left(r^{4}-4J^2\cos^2\theta\right)^{\frac{1}{2}}}\left(\dif t+{\frac {2J \sin^2\theta \,
 \dif\varphi}{r-M}}\right)^2\cr
&\quad+ \left(r^{4}-4J^2\cos^2\theta\right)^{\frac{1}{2}}\left[ {\frac {{\dif r}^{2}}{{(r-M)}^{2}}}+{\dif\theta}^{2}+
\sin^2 \theta\,{\dif\varphi}^{2} \right],\\
A_0&=-\sqrt{2}\,\frac{Mr-2J\cos\theta}{r^{2}-2J\cos\theta}\,,\qquad
A_\varphi=-\sqrt{2}\left(M \cos\theta
 -\frac{2Jr\sin^2 \theta}{r^{2}-2J\cos\theta}
 \right),\\
\phi&=\frac{\sqrt3}{2}\ln\left(\frac{r^{2}+2J\cos\theta}{r^{2}-2J\cos\theta}\right).
\end{align}
\end{subequations}
To our knowledge, the four-dimensional metric (\ref{single black hole metric}) first appeared in \cite{Rasheed:1995}. When $J=0$, the dilaton vanishes and this solution reduces to the $P=Q$ extremal Reissner--Nordstr\"om solution.

\section{Multi-centered rotating black hole solution}
\label{Rotating multi black hole sec}

Since the under-rotating extremal black hole solution can be written in the form (\ref{clement}), it is possible to generalise it to a multi-centered solution. For simplicity, we only consider the case of equal electric and magnetic charges, where the $A$ matrix is given by (\ref{A}). Using Cartesian coordinates ${\boldsymbol x}=(x,y,z)$, we can simply replace the harmonic functions (\ref{harmonic}) by
\begin{align}
\label{multi harmonic}
f=\sum_{n=1}^N\frac{M_n}{|{\boldsymbol x}-{\boldsymbol x}_n|}\,,\qquad g=\sum_{n=1}^N\frac{J_n(z-z_n)}{|{\boldsymbol x}-{\boldsymbol x}_n|^3}\,.
\end{align}
The corresponding solution then describes a superposition of $N$ black holes located at ${\boldsymbol x}={\boldsymbol x}_n$, each with mass $M_n$, angular momentum $J_n$, and charges $P_n=Q_n=\frac{M_n}{\sqrt2}$. Without loss of generality, we assume that all the black holes have different positions ${\boldsymbol x}_n$. Note that the spin vectors of the black holes are parallel or anti-parallel, in the sense that they point in either the $+z$ or $-z$ direction, depending on the sign of $J_n$. When $J_n=0$, this solution reduces to the Majumdar--Papapetrou solution with $P_n=Q_n$.\footnote{The five-dimensional form of this solution was studied in \cite{Matsuno:2012a}.}

The $\chi$ matrix for this solution can then be computed using (\ref{clement}), and we obtain 
\begin{align}
\chi=\begin{pmatrix}
-(1-f)^2-2g & -\sqrt{2}[(1-f)f-2g] & -f^2-2g \\
-\sqrt{2}[(1-f)f-2g] & -2f^2-4g+1 & ~\sqrt{2}[(1+f)f+2g] \\
-f^2-2g & ~\sqrt{2}[(1+f)f+2g] & -(1+f)^2-2g
\end{pmatrix}.
\end{align}
Using (\ref{chi}), we read off that $\tau=[(1+f)^2+2g]^{-1}$,
\begin{subequations}
\begin{align}
V_0&=-\frac{f^2+2g}{(1+f)^2+2g}\,,\\
V_5&=\sqrt{2}\,\frac{(1+f)f+2g}{(1+f)^2+2g}\,,
\end{align}
\end{subequations}and
\begin{subequations}
\begin{align}
\lambda_{00}&=\frac{2f^2-4g-1}{(1+f)^2+2g}\,,\\
\lambda_{05}&=-\sqrt{2}\,\frac{(1+f)f-2g}{(1+f)^2+2g}\,,\\
\lambda_{55}&=\frac{(1+f)^2-2g}{(1+f)^2+2g}\,.
\end{align}
\end{subequations}
With these expressions for $\tau$, $V_a$ and $\lambda_{ab}$, it can be checked that (\ref{twist}) is equivalent to the two equations (written in 3-vector notation):
\begin{align}
\label{two equations}
\nabla\times{\boldsymbol\omega}^0=-2\,\nabla g\,,\qquad
\nabla\times{\boldsymbol\omega}^5=-\sqrt{2}\,\nabla(f+2g)\,.
\end{align}
If we denote ${\boldsymbol{\tilde\omega}}^5\equiv{\boldsymbol\omega}^5-\sqrt{2}\,{\boldsymbol\omega}^0$, the solutions are given by\footnote{Each solution is defined only up to the gradient of a scalar, but this term can be set to zero by an appropriate transformation of $t$ and/or $x^5$ in (\ref{5D solution}).}
\begin{subequations}
\label{omegas}
\begin{align}
\label{omega0}
{\boldsymbol\omega}^0\cdot\dif{\boldsymbol x}&=-\sum_{n=1}^N\frac{2J_n}{|{\boldsymbol x}-{\boldsymbol x}_n|^3}[(y-y_n)\dif x-(x-x_n)\dif y]\,,\\
{\boldsymbol{\tilde\omega}}^5\cdot\dif{\boldsymbol x}&=\sqrt{2}\,\sum_{n=1}^N\frac{M_n(z-z_n)}{|{\boldsymbol x}-{\boldsymbol x}_n|}\,\frac{(y-y_n)\dif x-(x-x_n)\dif y}{(x-x_n)^2+(y-y_n)^2}\,.
\end{align}
\end{subequations}

Substituting the above expressions for $\tau$ and $\lambda_{ab}$ into (\ref{5D}), and after some rearrangement of the terms, we finally arrive at the five-dimensional metric:
\begin{align}
\label{5D solution}
\dif s_{(5)}^2&=\frac{H_-}{H_+}\left[\dif x^5-\sqrt{2}\,\dif t+\sqrt{2}\,\frac{1+f}{H_-}\left(\dif t+{\boldsymbol\omega}^0\cdot\dif{\boldsymbol x}\right)+{\boldsymbol{\tilde\omega}}^5\cdot\dif{\boldsymbol x}\right]^2\cr
&\quad-\frac{1}{H_-}\left(\dif t+{\boldsymbol\omega^0}\cdot\dif{\boldsymbol x}\right)^2+H_+\,\dif{\boldsymbol x}\cdot\dif{\boldsymbol x}\,,
\end{align}
where $H_\pm\equiv(1+f)^2\pm2g$, and ${\boldsymbol\omega}^0$, ${\boldsymbol{\tilde\omega}}^5$ are given by (\ref{omegas}). The four-dimensional form of this solution, after substituting in for $H_\pm$, is given by
\begin{subequations}
\label{multi black hole solution}
\begin{align}
\label{multi black hole metric}
\dif s_{(4)}^2&=-\frac{1}{\left[(1+f)^4-4g^2\right]^{\frac{1}{2}}}\left(\dif t+{\boldsymbol\omega^0}\cdot\dif{\boldsymbol x}\right)^2+\left[(1+f)^4-4g^2\right]^{\frac{1}{2}}\dif{\boldsymbol x}\cdot\dif{\boldsymbol x}\,,\\
\label{multi black hole gauge fields}
A_0&=-\sqrt{2}\,\frac{(1+f)f-2g}{(1+f)^2-2g}\,,\qquad A_i=\sqrt{2}\,\frac{1+f}{(1+f)^2-2g}\,\omega^0_i+\tilde\omega^5_i,\\
\label{multi black hole dilaton field}
\phi&=\frac{\sqrt3}{2}\ln\left[\frac{(1+f)^2+2g}{(1+f)^2-2g}\right].
\end{align}
\end{subequations}

\section{Properties of the multi-centered rotating black hole solution}
\label{Properties sec}

We begin by briefly recalling some properties of the single black hole solution (\ref{single black hole solution}) \cite{Rasheed:1995,Larsen:1999,Larsen:2000}. It is an asymptotically flat solution, with mass $M$ and angular momentum $J$. The solution also carries equal magnetic and electric charges, given by $P=Q=\frac{M}{\sqrt2}$. The parameters $M$ and $J$ are subject to the constraint $|J|<\frac{M^2}{2}$. When $M=J=0$, the solution reduces to empty Minkowski space-time.  

There is a degenerate event horizon located at $r=M$, with non-vanishing area $4\pi\sqrt{M^4-4J^2}$. There is also a curvature singularity at $r=\sqrt{2|J\cos\theta|}$, but it will always lie below the event horizon. These properties are consistent with the interpretation of the solution as an extremal black hole. However, we note that the event horizon has zero angular velocity, even if the black hole itself has a non-zero angular momentum. Moreover, there is no ergoregion present in the space-time. These somewhat unusual features are common to all the under-rotating extremal solutions.

Equipped with an understanding of the single black hole solution, we now turn to the multi-centered black hole solution (\ref{multi black hole solution}). From the form of the harmonic functions $f$ and $g$ in (\ref{multi harmonic}), we see that it is an asymptotically flat solution consisting of a superposition of $N$ separate sources in Minkowski space-time, each with mass $M_n$, angular momentum $J_n$, and charges $P_n=Q_n=\frac{M_n}{\sqrt2}$. To ensure that each source does indeed have a consistent interpretation as a black hole, we require the condition 
\begin{align}
\label{individual condition}
|J_n|<\frac{M_n^2}{2} 
\end{align}
to hold individually for $n=1,\dots,N$.

The event horizon of the $n$-th black hole is located at ${\boldsymbol x}={\boldsymbol x}_n$. This can be verified by introducing standard spherical polar coordinates $(r,\theta,\varphi)$ centered at ${\boldsymbol x}={\boldsymbol x}_n$, and taking the limit $t\rightarrow\frac{t}{\epsilon}$, $r\rightarrow\epsilon r$ and $\epsilon\rightarrow0$ of the metric (\ref{multi black hole metric}). Indeed, we recover the same near-horizon limit as that of the single black hole metric (\ref{single black hole metric}). It follows that the event horizon at $r=0$ is regular and has a non-vanishing area equal to $4\pi\sqrt{M_n^4-4J_n^2}$.

By inspection of the metric (\ref{multi black hole metric}), the only other points at which singularities can occur are where $(1+f)^2=2|g|$. By calculating the curvature invariants, it can be seen that they are curvature singularities. We now show that these curvature singularities all lie below the event horizons, in regions of the space-time not covered by the Cartesian coordinates used here. We first note that, for the $n$-th black hole,
\begin{align}
f_n^2-2|g_n|&=\frac{M_n^2}{|{\boldsymbol x}-{\boldsymbol x}_n|^2}-\frac{2|J_n||z-z_n|}{|{\boldsymbol x}-{\boldsymbol x}_n|^3}\cr
&\geq\frac{M_n^2-2|J_n|}{|{\boldsymbol x}-{\boldsymbol x}_n|^2}\cr
&>0\,,
\end{align}
where we have used the fact that $|z-z_n|\leq|{\boldsymbol x}-{\boldsymbol x}_n|$ and (\ref{individual condition}). It follows that
\begin{align}
(1+f)^2-2|g|&=\bigg(1+\sum_n f_n\bigg)^2-2\bigg|\sum_n g_n\bigg|\cr
&>1+\sum_n f_n^2-2\sum_n|g_n|\cr
&>1\,.
\end{align}
In the second step, we have used the fact that $f_n>0$ (since masses are always positive) as well as the triangle inequality. Hence $(1+f)^2>2|g|$ at every point on and outside the event horizons.

The existence of closed time-like curves (CTCs) in the space-time (\ref{multi black hole metric}) can also be ruled out, if it can be shown that the coordinates $(x,y,z)$ remain space-like everywhere in the space-time. Since ${\boldsymbol\omega^0}$ has both $x$ and $y$ components, this result will follow if $\left(\begin{smallmatrix}g_{xx}&g_{xy}\cr g_{xy}&g_{yy}\end{smallmatrix}\right)$ is a positive-definite matrix. This is equivalent to the positivity of its trace and determinant. Explicitly, these two conditions are
\begin{subequations}
\label{conditions}
\begin{align}
\label{positivity trace}
g_{xx}+g_{yy}=\frac{2(1+f)^4-8g^2-\omega^2-\bar\omega^2}{\left[(1+f)^4-4g^2\right]^{\frac{1}{2}}}&>0\,,\\
\label{positivity det}
g_{xx}g_{yy}-g_{xy}^2=(1+f)^4-4g^2-\omega^2-\bar\omega^2&>0\,,
\end{align}
\end{subequations}
where we have defined $\omega\equiv\omega^0_x$ and $\bar\omega\equiv\omega^0_y$ for brevity. Their expressions can be read off from (\ref{omega0}). 

It is straightforward to see that the second condition in (\ref{conditions}) is stronger than the first. To show that the second condition is satisfied, we first note that, for the $n$-th black hole,
\begin{align}
f_n^4-4g_n^2-\omega_n^2-\bar\omega_n^2&=\frac{M_n^4}{|{\boldsymbol x}-{\boldsymbol x}_n|^4}-\frac{4J_n^2\left[(x-x_n)^2+(y-y_n)^2+(z-z_n)^2\right]}{|{\boldsymbol x}-{\boldsymbol x}_n|^6}\cr
&=\frac{M_n^4-4J_n^2}{|{\boldsymbol x}-{\boldsymbol x}_n|^4}\cr
&>0\,,
\end{align}
where we have used (\ref{individual condition}) in the last step. It follows that
\begin{eqnarray}
\label{inequality 1}
(1+f)^4-4g^2-\omega^2-\bar\omega^2&=&\bigg(1+\sum_n f_n\bigg)^4-4\bigg(\sum_n g_n\bigg)^2-\bigg(\sum_n\omega_n\bigg)^2-\bigg(\sum_n\bar\omega_n\bigg)^2\cr
&>&1+\sum_n\left(f_n^4-4g_n^2-\omega_n^2-\bar\omega_n^2\right)\cr
&&\hskip0.13in+\sum_{n\neq m}\left(3f_n^2f_m^2-4g_ng_m-\omega_n\omega_m-\bar\omega_n\bar\omega_m\right) \cr
&>&1+\sum_{n\neq m}\left(3f_n^2f_m^2-4g_ng_m-\omega_n\omega_m-\bar\omega_n\bar\omega_m\right).
\end{eqnarray}
In the second step, we have used the fact that $f_n>0$. But
\begin{align}
\label{inequality 2}
3f_n^2f_m^2-4g_ng_m-\omega_n\omega_m-\bar\omega_n\bar\omega_m
&\geq3f_n^2f_m^2-4|g_n||g_m|-|\omega_n||\omega_m|-|\bar\omega_n||\bar\omega_m|\cr
&=\frac{3M_n^2M_m^2}{|{\boldsymbol x}-{\boldsymbol x}_n|^2|{\boldsymbol x}-{\boldsymbol x}_m|^2}-\frac{4|J_n||J_m|}{|{\boldsymbol x}-{\boldsymbol x}_n|^3|{\boldsymbol x}-{\boldsymbol x}_m|^3}\times\cr
&\quad\left(|x-x_n||x-x_m|+|y-y_n||y-y_m|+|z-z_n||z-z_m|\right)\cr
&\geq\frac{3M_n^2M_m^2-12|J_n||J_m|}{|{\boldsymbol x}-{\boldsymbol x}_n|^2|{\boldsymbol x}-{\boldsymbol x}_m|^2}\cr
&>0\,,
\end{align}
where we have used the fact that $|x-x_n|,|y-y_n|,|z-z_n|\leq|{\boldsymbol x}-{\boldsymbol x}_n|$ in the third step. We have also used (\ref{individual condition}) in the last step. Putting (\ref{inequality 1}) and (\ref{inequality 2}) together, we see that (\ref{positivity det}) is satisfied. This implies that (\ref{positivity trace}) is also satisfied, so the space-time (\ref{multi black hole metric}) does not contain any CTCs.

We thus conclude that the superposition of multiple rotating sources in (\ref{multi harmonic}) does not lead to any new pathologies such as naked singularities or CTCs. The fact that the metric is smooth and invertible outside the event horizons means that the space-time does not contain other defects such as conical singularities or Dirac--Misner strings \cite{Hartle:1972}. The interpretation of the solution (\ref{multi black hole solution}) as a balanced superposition of $N$ rotating black holes thus appears to be a consistent one.

\section{Solution for two black holes along a symmetry axis}
\label{Special case sec}

The general solution (\ref{multi black hole solution}) will simplify considerably when the black holes are all placed along a straight line in the $z$-direction. This is due to the solution becoming rotationally symmetric about this axis. In this section, we present the explicit form of this solution for the case of two black holes. Without loss of generality, we choose them to be located at $(x_1,y_1,z_1)=(0,0,-\alpha)$ and $(x_2,y_2,z_2)=(0,0,\alpha)$, respectively, where $\alpha$ is a positive constant. If we assume that the black holes have masses $M_1$ and $M_2$, and angular momenta $J_1$ and $J_2$, respectively, the corresponding harmonic functions are
\begin{align}
f=\frac{M_1}{R_1}+\frac{M_2}{R_2}\,,\qquad
g=\frac{J_1(z+\alpha)}{R_1^3}+\frac{J_2(z-\alpha)}{R_2^3}\,,
\end{align}
where $R_1\equiv\sqrt{x^2+y^2+(z+\alpha)^2}$ and $R_2\equiv\sqrt{x^2+y^2+(z-\alpha)^2}$.

It is natural to introduce polar coordinates $(\rho,\varphi)$ given by
\begin{align}
\label{polar coords}
x=\rho\cos\varphi\,,\qquad y=\rho\sin\varphi\,.
\end{align}
The coordinates $(t,\rho,z,\varphi)$ are then the standard Weyl--Papapetrou coordinates used for describing stationary and axisymmetric space-times. In fact, since there are only two black holes in this case, it is possible to further transform $(\rho,z)$ to prolate spheroidal coordinates $(\xi,\eta)$ given by
\begin{align}
\rho=\alpha\sqrt{(\xi^2-1)(1-\eta^2)}\,,\qquad z=\alpha\xi\eta\,.
\end{align}
These new coordinates take the ranges $1\leq\xi<\infty$ and $-1\leq \eta\leq1$. The two black holes are then located at $(\xi,\eta)=(1,-1)$ and $(\xi,\eta)=(1,1)$, respectively. They divide the $z$-axis into three parts: the inner part between the black holes is given by $\xi=1$, while the two outer parts that extend to infinity are given by $\eta=\pm1$. Asymptotic infinity itself is reached when $\xi\rightarrow\infty$ while keeping $\eta$ fixed. 

It can be checked that, in prolate spheroidal coordinates, the four-dimensional metric (\ref{multi black hole metric}) is given by
\begin{align}
\dif s_{(4)}^2&=-\frac{\alpha^2(\xi^2-\eta^2)^3}{(H_+H_-)^{\frac{1}{2}}}\left(\dif t+\omega_\varphi\,\dif\varphi\right)^2\cr
&\quad+\frac{(H_+H_-)^{\frac{1}{2}}}{(\xi^2-\eta^2)^2}\,\bigg\{\frac{\dif\xi^2}{\xi^2-1}+\frac{\dif\eta^2}{1-\eta^2}+\frac{(\xi^2-1)(1-\eta^2)\,\dif\varphi^2}{\xi^2-\eta^2}\bigg\}\,,
\end{align}
where
\begin{subequations}
\begin{align}
\omega_\varphi&\equiv{\frac { 2( {\xi}^{2}-1) ( 1-\eta^{2})
 \left[  ( \xi-\eta ) ^{3}J_1+ ( \xi+\eta) ^{3} 
J_2 \right] }{\alpha ( {\xi}^{2}-\eta^{2} ) ^{3}}}\,,\\
H_\pm&\equiv( {\xi}^{2}-\eta^{2} )  \left[ \alpha ( {\xi}^{2}-\eta^{2} ) + ( \xi-\eta )M_1 + ( \xi+\eta
)M_2  \right] ^{2}\cr
&\quad\pm2 \left[( \xi-\eta ) ^{3
}( 1+\xi\eta )  J_1-( \xi+\eta) ^{3} ( 1-\xi\eta) J_2\right].
\end{align}
\end{subequations}
The gauge fields (\ref{multi black hole gauge fields}) and dilaton field (\ref{multi black hole dilaton field}) are given by
\begin{subequations}
\begin{align}
A_0&=\frac{\sqrt{2}}{H_-}\Big\{ -( {\xi}^{2}-\eta^{2})  \left[  ( \xi-\eta
 )M_1 +( \xi+\eta) M_2  \right] \cr
&\hskip0.6in \times
\left[ \alpha
 ( {\xi}^{2}-\eta^{2}) + ( \xi-\eta)M_1 + 
 ( \xi+\eta)M_2  \right] \cr
&\hskip0.6in+2 \left[( \xi-\eta) ^{3}( 1+\xi\eta)J_1- ( \xi+\eta) ^{3} ( 1-\xi\eta) 
J_2\right]
\Big\}\,,\\
A_\varphi&=\frac{\sqrt{2}}{H_-}\Big\{-\left[  ( \xi-\eta ) ( 1+\xi\eta )  M_1- ( \xi+\eta
 )  ( 1-\xi\eta ) M_2 \right] \cr
&\hskip0.6in \times\left[ \alpha
 ( {\xi}^{2}-\eta^{2} ) +( \xi-\eta )M_1 +( \xi+\eta )M_2  \right] ^{2}\cr
&\hskip0.6in+2( \xi-\eta ) ^{2}
 \big[  ( {\xi}^{2}-\eta^{2} ) M_1+ ( {\xi}^{2}-2+\eta
^{2} ) M_2 \big] J_1\cr
&\hskip0.6in+2( \xi+\eta ) ^{2}
 \big[  ( {\xi}^{2}-2+\eta^{2} ) M_1+ ( {\xi}^{2}-\eta
^{2} ) M_2 \big] J_2\cr
&\hskip0.6in+2\alpha( {\xi}^{2}-1 )( 1-\eta^{2} ) 
   \left[  ( \xi-\eta ) ^{3}J_1+
 ( \xi+\eta ) ^{3}J_2 \right] \Big\}\,,\\
 \phi&=\frac{\sqrt3}{2}\ln\left(\frac{H_+}{H_-}\right).
\end{align}
\end{subequations}
The advantage of using these coordinates is that the square-root functions $R_1$ and $R_2$ have become algebraic functions of $\xi$ and $\eta$. This would make the solution more convenient to study than in the original coordinates.

\section{Conclusion}

In this paper, we have presented a solution of Kaluza--Klein theory describing an arbitrary superposition of extremally charged, rotating black holes. Each black hole carries equal electric and magnetic charge, and has an angular momentum that is allowed to lie in a certain range. The spin vectors of the black holes are parallel or anti-parallel, in the sense that they point in the same or opposite directions.

This solution is specified by two harmonic functions on a three-dimensional flat base space, which have poles at the locations of the black holes. By choosing the harmonic functions appropriately, we can obtain interesting special cases of this solution. We have considered one special case in this paper, namely, when the two black holes are placed along the $z$-axis, so that the solution has a rotational symmetry. An even more interesting example would be if the two black holes are not placed along the $z$-axis, but say along the $x$-axis. Such a solution will not have any spatial symmetry, and so will be considerably more complicated than the previous example. 

The fact that the black holes in our solution are in static equilibrium means that there is an exact cancellation of the forces between them. When the black holes are non-rotating, this cancellation of forces is well understood in the limit of large separation: the electrostatic and magnetostatic repulsion between them balances out their Newtonian gravitational attraction. But when the black holes are rotating, they each acquire electric and magnetic dipole moments, thus complicating the interactions between them. Besides the spin-spin and dipole-dipole interactions (familiar from the Israel--Wilson--Perj\'es solution \cite{Kastor:1998}), there are also dipole-monopole interactions to consider. Moreover, there will be interactions arising from the dilaton field. It would be interesting to understand how the various forces balance out in this case.

One can also consider possible generalisations of the solution found in this paper. For simplicity, we have only considered the case of equal electric and magnetic charges. It should be possible to extend the solution to the more general case of unequal electric and magnetic charges, with the ratio between them being the same for each black hole. A more intriguing question is whether the condition of parallel or anti-parallel black hole spins can be relaxed. This would require one to consider a more general form for the harmonic function $g$ than in (\ref{multi harmonic}), and find the corresponding set of solutions to (\ref{two equations}). We leave these interesting questions for future work.

\section*{Acknowledgements}

ET wishes to thank Brenda Chng, for her collaboration at the early stages of this work. He also wishes to thank Yu Chen and Kenneth Hong for helpful discussions.

\end{document}